# Simulating galaxy formation

M Steinmetz

Max–Planck–Institut für Astrophysik, Karl–Schwarzschild–Straße 1, Postfach 1523, 85740 Garching, Germany

**Abstract.** A review on numerical simulations of galaxy formation is given. Different numerical methods to solve collisionless and gas dynamical systems are outlined and one particular simulation technique, *Smoothed Particle Hydrodynamics* is discussed in some detail. After a short discussion of the most relevant physical processes which affect the dynamics of the gas, the success and shortcomings of state of the art simulations are discussed via the example of the formation of disk galaxies.

## 1. Introduction

Within the last twenty years, our understanding has greatly improved of how galaxies form and how their distribution on large scales can be explained. From the theoretical point of view, we have benefited substantially from numerical $N$–body–simulations. These simulations, however, only follow the evolution of the distribution of dark matter, which is commonly believed to dominate the dynamics on large scales ($\gtrsim 10\,\mathrm{Mpc}$), but which, by definition, cannot be observed directly. During the last few years it has become possible to extend $N$–body simulations to include also gas dynamical effects. Pioneering work was done by Evrard (1988), Hernquist & Katz (1989), Navarro & Benz (1991), Cen (1992) and Umemura (1993). These simulations allowed a more realistic description of the dynamics on small scales ($< 5\,\mathrm{Mpc}$), which is affected or even dominated by gas dynamical effects. Moreover, on larger scales combined gas dynamical and $N$–body simulations provide a link between dark matter and directly observable quantities. Today, gas–dynamical simulations are being applied to a large variety of problems in galaxy formation and large scale structure modelling. Without trying to be complete, I can to mention galaxy formation (Katz & Gunn 92, Navarro & White 1994, Steinmetz & Müller 1994, 1995, Evrard, Summers & Davis 1994), galaxy biasing (Katz, Hernquist & Weinberg 1992, Cen & Ostriker 1992), X–ray Clusters (Evrard 1990, Thomas & Couchman 1992, Katz & White 1993, Bryan *et al* 1994, Navarro, Frenk & White 1995b), and Ly$_\alpha$ absorption systems (Hernquist *et al* 1995, Miralda-Escude *et al* 1995, Petitjean, Mücket & Kates 1995). Complementary to these three dimensional studies, one dimensional simulations which exhibit a much higher resolution have also been performed (Thoul & Weinberg 1995a).

In this lecture, I would like to give a compact but comprehensive overview of one particular hydrodynamical simulation technique, *Smoothed Particle Hydrodynamics* (SPH). This article is, therefore, complementary to the lectures by Klypin in this volume, which concentrate on grid based hydro methods. In the next section, I summarize different numerical techniques to perform N–body–simulations. As an alternative to these software technique I also give an introduction to a completely different approach, namely the use of special purpose hardware. Section 3 deals with hydrodynamics, section 4 with related physical processes like radiation cooling and heating due to feedback processes. In section 5 the use of such hydro methods is illustrated for the case of the formation of galaxies. This part is again complementary to the lectures by Klypin and by Primack, which mainly focus on the large scale distribution of galaxies.

## 2. N–body–simulations

It is widely believed, that the dominant fraction of the mass of the universe consists of non–baryonic dark matter. Attractive candidates have been postulated by particle physics, namely a weakly interacting elementary particle with a mass of several hundreds of GeV (see also the lectures on particle physics in this volume), but also more massive astrophysical objects like brown dwarfs or black holes are under discussion (see, e.g. Griest, this volume). These particles/objects of mass $m$ interact only gravitationally and their distribution function $f(\mathbf{r}, \mathbf{v}, t)$ can mathematically be described by the Vlasov equation (collisionless Boltzmann equation, Binney & Tremaine 1987):

$$\frac{df(\mathbf{x}, \mathbf{v}, t)}{dt} = \frac{\partial f}{\partial t} + \mathbf{v} \cdot \nabla f - \nabla \Phi \frac{\partial f}{\partial \mathbf{v}} = 0 \tag{1}$$

The gravitational potential is given by Poisson's equation

$$\Delta \Phi = 4\pi G \int d^3 v \, m \, f(\mathbf{x}, \mathbf{v}, t) \,. \tag{2}$$

It turns out, that $N$–body simulations provide a robust and efficient, although sometimes computationally expensive tool to solve equations (1-2) numerically (Efstathiou *et al* 1985). In $N$–body–simulations, the trajectories of particles are determined by the laws of Newtonian dynamics

$$\frac{d\mathbf{v}_i}{dt} = -\nabla \Phi|_i \tag{3}$$

$$\Phi(\mathbf{r}_i) = -G \sum \frac{m_j}{|\mathbf{r}_j - \mathbf{r}_i|} \tag{4}$$

Every particle of a $N$–body–simulation represents a huge number of dark matter particles (compare the mass of a body in the simulation (typically $10^8 - 10^{12}$ M$_\odot$ with that of an elementary particle (100 GeV)!). $N$–body–simulations can, therefore, be interpreted as a Monte–Carlo–Approximation of the Vlasov equation, i.e. the set $(\mathbf{r}_i(t), \mathbf{v}_i(t), i = 1, N)$ samples the distribution function $f(\mathbf{r}, \mathbf{v}, t)$.

### 2.1. Numerical methods to solve the $N$–body System

The main computational problem concerning the integration of (3-4) is posed by the long range nature of gravity. Every particle interacts with every other one and the interaction is decaying too slowly with distance. Consequently, the computational effort grows with the particle number $N$ like $N^2$ which makes it prohibitively expensive to integrate systems larger than a few $10^4$ particles directly, even on the largest supercomputer available today. The direct summation approach to solve equation (4) is sometimes referred to as the *PP* (= Particle Particle) method. Several techniques were developed to allow integration of larger systems (see also Hockney & Eastwood 1988). They can roughly be subdivided into two classes.

(i) Particle based methods try to reduce the number of pairwise interactions. For this purpose, more distant particles are systematically grouped together and the combined force of all the members of such a group is replaced by a multipole expansion of the force exerted by the group. One can show that in order to calculate the force on one particle, the $N$ interactions of the PP method can be replaced by $\propto \log N$ multipole forces. The total computing time to perform a time step is $\propto N \log N$. In order to systematically group particles together and in order to set up $\log N$ multipole forces, a tree data structure is commonly used (Apple 1981, Porter 1985, Barnes & Hut 1986). Therefore, these codes are called *treecodes*. A related method, which is is often used in molecular dynamics, is the *fast multipole method* (FFM, Rokhlin & Greengard 1987), where the CPU time only scales like $N$.

(ii) Mesh based methods (*PM* = Particle Mesh) solve (4) using Poisson's equation (2). At first, particle positions are assigned to a grid, the number of particles per mesh cell defining a density field $\varrho(\mathbf{r})$. By means of Fast–Fourier–Transform (FFT) (or by another mesh based Poisson solver) Poisson's equation is solved and the gravitational potential $\Phi$ is obtained at the vertices of the mesh. By interpolation between the mesh vertices, the force at the position of any particle can be calculated. The CPU time for PM methods scales linearly with the particle number $N$.

Both approaches have their disadvantages and their merits and the optimal choice is problem dependent. The main advantage of the mesh based methods is that they are quite fast. Periodic boundary conditions, which are usually applied for cosmological simulations, are automatically implemented. The disadvantage is that the dynamical range is rather limited. Current parallel supercomputers can handle meshes with up to $1500^3$ cells, while on a large memory workstation one can typically handle a $256^3$ mesh. The performance of PM codes is limited by the available memory rather than the CPU time. Since the mesh force is quite inaccurate for distances less than twice the size of a cell, the dynamic range (the ratio box-size to resolution scale) is at best several hundreds, while the numerically more expensive tree codes can easily handle a dynamic range above one thousand, and even above ten thousand. Although the dynamic range is large, the more expensive computing time of treecodes does not allow such large particle numbers as in PM. Furthermore, periodic boundaries are not automatically provided but have to be added explicitely, e.g., by means of the Ewald (1921) summation technique (Hernquist, Bouchet & Suto 1991). A widely used compromise between mesh and particle based methods is the hybrid scheme P3M (Particle-Particle–Particle-Mesh). The long range force is calculated by PM, but the force between particles having a distance less than twice the size of a mesh cell is modified using PP. Therefore, a good compromise between the dynamic range of particle methods and the speed of mesh based methods can be obtained. However, in the case of highly clustered systems (as an extreme example consider that all particles are within one mesh cell), the performance is degraded to less than that of a PP algorithm. These drawback can be avoided by means of adaptive submeshes (Couchman 1991, 1995): Whenever too many particles are found within a region of the grid, a finer grained submesh is used in this region and an additional PM step is performed. PP is only applied below the finest refined grid.

2.2. *The special purpose hardware GRAPE*

A completely different approach to solve the $N$–body system was invented by Sugimoto *et al* (1990) at the University of Tokyo. A closer look at equation (4) shows that the calculation of a force pair involves only very few very specific operations. Furthermore, these operations are to a large part mutually independent, e.g., the difference $\Delta x$ can be calculated independently from $\Delta y$. These few operations can be hardwired in a specially designed chip. Such a highly specialized chip can, of course, reach much higher Flop rates than a conventional computer. A board containing 8 such chips, GRAPE (=Gravity PipE, see also figure 1), can nowadays be ordered from Sugimoto's group. The board is connected to a workstation (e.g., a SUN Sparc 20) via the VME interface and reaches a sustained performance of 5 Gflop/s per board, which can be further improved by using several boards together. Beside the calculation of the gravitational forces GRAPE also provides a list of neighboured particles, a feature which turns out to be quite convenient to adopt the hydrodynamical method SPH to GRAPE (Umemura *et al* 1993, Steinmetz 1996). Nevertheless, GRAPE is still based on a simple PP algorithm with a $N^2$ scaling of the CPU time. Therefore, for a sufficiently large particle number, GRAPE is always inferior to the software methods presented above. It is, therefore, predominantly suited to highly clustered systems like individual galaxies or galaxy clusters. The high spatial resolution which is required to investigate such systems also imposes a very small time step (see below). To run simulations which cover a Hubble time, several ten thousand time steps can be necessary, far more than usually used in cosmological $N$–body simulations (several hundreds up to a few thousands). Using the same amount of CPU time, such high resolution simulations have correspondingly fewer particles (typically fewer than several $10^5$) and reach a regime, where a PP code on GRAPE is faster than a treecode on a supercomputer. Moreover, these systems not only have a large range of involved length scales but also quite different time scales. By means of a individual time step scheme they can be accelerated by an order of magnitude. Such a multiple time step scheme is easily and efficiently implemented on GRAPE (Steinmetz 1996).

**Figure 1.** Block diagram of the force–calculation unit of GRAPE. Besides the calculation of the softened Newtonian force law also a list of neighboured particles is delivered (from Okumura *et al* 1993).

But also the regime of simulations with very large particle numbers is suitable to GRAPE. Makino & Funato (1993) modified a treecode to run on GRAPE. Brieu, Summers & Ostriker (1995) used GRAPE to speed up the PP part of P3M. They could reach a performance of 50% of a Cray C90. Both approaches, GRAPE–TREE and GRAPE–P3M, seem to provide a performance comparable to todays supercomputers, but the size of the simulations is mainly limited by the memory, CPU speed and input/output performance of the front end, a limitation which may be overcome within the next few years. Due to the low price (of a medium level work station) another big advantage of the GRAPE is that only a very few users share the CPU time, while in supercomputer centers one has to write proposals and seldom one can hope to get more than about 10% of a machine. After a proposal is granted, one has to compete with different other groups. The resulting ratio of turnaround time to actual CPU time is seldom better than 5, while on a GRAPE it is almost 1.

## 3. Hydrodynamics

In this section, I give a compact overview of a particle based method which numerically solves the hydrodynamic equations. It is usually known as *Smoothed Particle Hydrodynamics* (SPH) and was independently developed by Lucy (1977) and Gingold & Monaghan (1977, 1982). For a modern and detailed review we refer to Benz (1990) and Monaghan (1992). Details of different implementations developed to study the formation of galaxies can be found in Evrard (1988), Hernquist & Katz (1989), Katz, Weinberg & Hernquist (1995), Navarro & White (1993), Couchman, Thomas & Pearce (1995), Steinmetz & Müller (1993) and Steinmetz (1996).

### 3.1. The hydrodynamics equations

The evolution of a system of ions can be described by the Boltzmann equation, provided one includes collisions, i.e., $\frac{df}{dt} = [\frac{df}{dt}]_c$. In order to approximate such a system by the gas–dynamical equations, two requirements have to be fulfilled (see e.g., Müller 1994): The mean free path of an ion must be small compared to the typical scale of the object under consideration, i.e., the microscopic behaviour of the gas particles is negligible. Secondly, the forces between particles must saturate (which means they are short ranged), otherwise collective effects may be dominant. Formally, the saturation can be written as

$$\lim_{N \to \infty} \left( \frac{E}{N} \right) = \text{const.}, \tag{5}$$

$E/N$ being the energy per particle. An example for a non–saturating force is gravity. For a self gravitating system of bosons, $E/N \propto N^2$, while $E/N \propto N^{4/3}$ for fermions. Consequently, gravity has to be included as an external force in the hydrodynamical equations. Physically, the hydrodynamic equations reflect the conservation of mass, energy and momentum (see, e.g., Potter 1973, Chorin & Marsden 1979, Morawetz 1981). In the case of a continuous flow field (i.e. no shock waves or contact discontinuities) these conservation laws can be written as a set of partial differential equations. In the following I will restrict myself to the Lagrangian formulation, i.e. I assume coordinates which are comoving with the fluid element †. Mass, momentum and energy conservation can then be expressed as

$$\frac{d\varrho}{dt} = - \varrho \nabla \mathbf{v} \tag{6}$$

$$\frac{d\mathbf{v}}{dt} = - \frac{1}{\varrho} \nabla P - \nabla \Phi \tag{7}$$

$$\frac{d\varepsilon}{dt} = - \frac{P}{\varrho} \nabla \mathbf{v} + \frac{Q - \Lambda(\varrho, T)}{\varrho} , \tag{8}$$

where $\varrho$ is the density of the fluid, $\mathbf{v}$ its velocity, $P$ its pressure and $\varepsilon$ the specific thermal energy, respectively. $\Phi$ is the gravitational potential. Equation (7) is usually referred to as "Euler's equation". The energy equation (8) is derived from the first law of thermodynamics. $\Lambda$ and $Q$ denote energy sinks and sources (see below). Euler's equation (7) can also be derived from a least action principle (see e.g., Lamb 1879, Eckart 1960) The system is closed by Poisson's equation (2) and by an equation of state, which in the case of an ideal gas reads

$$P = (\gamma - 1)\,\varepsilon\,\varrho \tag{9}$$

with $\gamma = \frac{5}{3}$.

† Note that finite difference methods are usually formulated in Eulerian coordinates which are fixed in space (see, e.g., Klypin this volume).

### 3.2. Smoothed Particle Hydrodynamics

The basic concept of SPH is to replace a field $A(\mathbf{r}, t)$ with an estimate $A_s(\mathbf{r}, t)$ which is smoothed on a length scale $h$

$$A_s(\mathbf{r}, t) = \int d^3 r'\, A(\mathbf{r}', t)\, W(\mathbf{r} - \mathbf{r}', h)\,. \tag{10}$$

The kernel function $W$, which is strongly peaked at $r = 0$ (e.g. a Gaussian), is required to converge to a $\delta$-distribution in the limit $h \to 0$. Furthermore, the smoothing procedure commutes with derivatives. Hence one obtains, e.g.,

$$\begin{aligned}
\nabla A_s &= \int d^3 r'\, A(\mathbf{r}', t)\, \nabla W(\mathbf{r} - \mathbf{r}', h) \\
&= -\int d^3 r'\, A(\mathbf{r}', t)\, \nabla' W(\mathbf{r} - \mathbf{r}', h) \\
&= \int d^3 r'\, \nabla'[A(\mathbf{r}', t)\, W(\mathbf{r} - \mathbf{r}', h)] = (\nabla A)_s\,,
\end{aligned} \tag{11}$$

where $\nabla$ and $\nabla'$ are the nabla operators with respect to $\mathbf{r}$ and $\mathbf{r}'$. In the third step of the last equation we integrated by parts and used that $W$ is sufficiently rapidly decaying for $r \to \infty$, i.e., the surface term vanishes.

So far we have only defined a smoothing procedure. Now we make two approximations:

(i) A physical quantity $A$ is identified with its smoothed estimate $A_s$. We, therefore, drop the subscript s in the following.

(ii) The field $A$ is discretized by considering it at a number of $N$ positions $\mathbf{r}_i, i = 1, \ldots, N$. These marker points are assigned with the mass of the fluid element $m_i$ and, according to our choice of Lagrangian variables, they are moving with the fluid. With $A_i = A(\mathbf{r}_i)$, we can approximate (10) by a finite sum. For the density, we obtain for example

$$\varrho(\mathbf{r}_i) = \sum_{j=1}^{N} m_j\, W(\mathbf{r}_i - \mathbf{r}_j, h)\,. \tag{12}$$

and for an arbitrary field $A$

$$A(\mathbf{r}_i) = \sum_{j=1}^{N} m_j\, \frac{A_j}{\varrho_j}\, W(\mathbf{r}_i - \mathbf{r}_j, h)\,. \tag{13}$$

Note that (13) can be interpreted as a Monte–Carlo–integration of (10). However, the $\mathbf{r}_i$ are not a Poisson sample but they are sampled by the dynamics of the fluid itself. Therefore, the error the error scales $\propto N^{-1}$ and not $\propto N^{1/2}$ as one naively might expect. For a detailed discussion and an error analysis we refer to Monaghan (1992) and references therein.

The summation in (13) formally extends over all particles. However since $W$ is strongly peaked at $r = 0$ and since $W$ is at least exponentially decaying for $r \to \infty$, the sum only extends over nearby particles. From the computational point of view it is, therefore, essential for the performance of the program to find these neighbours efficiently. One possibility is, e.g., to use the interaction list of a tree code or to use the neighbour list as it is delivered by GRAPE.

An elegant way to derive the Euler equation in the SPH approximation is shown by Gingold & Monaghan (1982): the density (12) is put in the Lagrangian and the action is minimized for variations along stream lines. One then obtains

$$\frac{d\mathbf{r}_i}{dt} = -\sum_{j=1}^{N} m_j \left( \frac{P_i}{\varrho_i^2} + \frac{P_j}{\varrho_j^2} \right) \nabla_i W(\mathbf{r}_i - \mathbf{r}_j, h) - \nabla \Phi. \tag{14}$$

The gravitational acceleration $\nabla \Phi$ is evaluated as in the case of a N–body system. Note, that the pairwise force in (14) is antisymmetric in the indices $i$ and $j$, i.e., Newton's third law is strictly fulfilled. Therefore, momentum and angular momentum are exactly conserved (Monaghan 1992). The advantage of the derivation via a Lagrangian is that it automatically and uniquely provides a conservative formulation. Of course, one also could derive the SPH equations by applying the smoothing procedure directly to Euler's equation (7), but the resulting pairwise force would not be antisymmetric. By reordering differentiation and smoothing in a tricky way one also can derive formulations which are antisymmetric in $i$ and $j$, but there exists an infinite set of analytically equivalent formulations † (see e.g. Monaghan 1992). My personal preference is (14), since it (i) arises naturally from a least action principle and (ii) it exhibits the best and most stable results in the test problems I have performed so far. To my knowledge, it is also the most widely used formulation of the equation of motion.

Similar to Euler's equation, the energy equation can be derived, too by applying the SPH smoothing procedure to (8), Again the same (formal) problem exists thay on can derive a infinite set of analytically equivalent formulations. However, since we already fixed the representation of (7), the energy equation should be consistently formulated, i.e. the change in total kinetic energy $E_{\text{kin}}$ should exactly compensate the change in thermal energy $E_{\text{th}}$ (for $\Lambda, Q = 0$), i.e.

$$\begin{aligned}\frac{dE_{\text{kin}}}{dt} &= \frac{d}{dt}\left( \frac{1}{2} \sum m_i \mathbf{v}_i^2 \right) = \sum m_i \mathbf{v}_i \frac{d\mathbf{v}_i}{dt} \\ &= -\frac{d}{dt}\left( \sum m_i \varepsilon_i \right) = -\frac{dE_{\text{th}}}{dt}.\end{aligned} \tag{15}$$

Inserting (14) into the upper equation, one obtains after some algebra (Benz 1990)

$$\frac{d\varepsilon_i}{dt} = \frac{P_i}{\varrho_i^2} \sum m_j (\mathbf{v}_i - \mathbf{v}_j) \nabla_i W(\mathbf{r}_i - \mathbf{r}_j, h) + \frac{Q - \Lambda}{\varrho}. \tag{16}$$

† The situation is similar in finite difference methods, where there also exists a large variety of different ways how to relate a quantity $A$ on the grid to a finite difference estimate of its gradient $\nabla A$.

Since in the adiabatic case $\frac{d}{dt}\varepsilon = \frac{P}{\varrho^2}\frac{d\varrho}{dt} = -\frac{P}{\varrho}\nabla\mathbf{v}$, we also obtain a consistent expression for the divergence of the velocity field:

$$(\nabla\mathbf{v})_i = -\frac{1}{\varrho_i}\sum m_j\,(\mathbf{v}_i - \mathbf{v}_j)\,\nabla_i W(\mathbf{r}_i - \mathbf{r}_j, h). \tag{17}$$

### 3.3. Shock waves

Shock waves, are a common phenomenon in astrophysics. During galaxy formation they can be quite extreme. Consider a Milky Way sized halo with a virial temperature of about $10^6$ K, i.e. a circular velocity of 200 km/sec. Due to gravity the gas in the halo will typically fall towards the galaxy with a comparable velocity. But the gas within the galaxy is much colder, namely about $10^4$ K, i.e. its sound velocity $c_s$ is of the order of 10 km/sec which implies Mach numbers of about 20.

Since shock waves are discontinuities in the flow, they cannot be described by differential equations (6-8) but only by the integral formulation of the conservation laws (see e.g. Morawetz 1981). This fact poses no problem to modern finite difference methods, which are generally derived from the integral formulation of the hydrodynamical equations (6-8). For SPH, which intrinsically assumes smooth fields, such an approach could not be successfully implemented so far. Instead, SPH uses an *artificial viscosity* similar to most of the older finite difference methods. It is known from hydrodynamics, that in fluids with non-vanishing viscosity, shock waves are smoothed over a finite length scale determined by the interaction length of the process which is responsible for the viscosity. Hence the flow can be described by the partial differential equation (6-8), modified by additional terms describing the influence of viscosity. In the limit of vanishing viscosity (or at distances far from the shock front), these viscous solutions converge to the discontinuous behaviour of an ideal fluid. In numerical hydrodynamics, one introduces an artificial viscosity, which mimics a physical viscosity in regions of strong compression. The viscosity coefficients are determined by the resolution scale. This type of viscosity acts such that the discontinuity is smoothed over a few resolution length scales to a steep, but finite gradient. Typically, a mixture of a bulk and a von–Neumann–Richtmyer viscosity (Monaghan & Gingold 1983) is applied. The artifical viscosity $Q$ can be written as

$$Q = -\frac{\alpha\,h\,c_s}{\varrho}\,\nabla\mathbf{v} + \frac{\beta\,h^2}{\varrho}\,(\nabla\mathbf{v})^2, \quad \text{if } \nabla\mathbf{v} < 0 \tag{18}$$

where $\alpha, \beta$ are free parameters of order unity. The equation of motion is modified according to

$$\left(\frac{P_i}{\varrho_i^2} + \frac{P_j}{\varrho_j^2}\right) \rightarrow \left(\frac{P_i}{\varrho_i^2} + \frac{P_j}{\varrho_j^2} + Q_{ij}\right). \tag{19}$$

with $Q_{ij} = \frac{1}{2}(Q_i + Q_j)$. Analogously, the energy equation has to be modified to guarantee energy conservation:

$$\frac{d\varepsilon_i}{dt} \rightarrow \frac{d\varepsilon_i}{dt} + \frac{1}{2}\sum m_i\,Q_{ij}(\mathbf{v}_i - \mathbf{v}_j)\cdot\nabla_i W(\mathbf{r}_i - \mathbf{r}_j, h) \tag{20}$$

The disadvantage of the viscosity is that the $\nabla\mathbf{v}$ term, which is determined by averaging over nearby particles, is not well localized, i.e., it is not sufficient to damp post shock oscillations. Monaghan & Gingold (1983), therefore, proposed an alternative viscosity tensor, where $\nabla \mathbf{v}$ is approximated by the interparticle velocity differences:

$$Q_{ij} = \begin{cases} \frac{-\alpha c_{ij}\mu_{ij} + \beta \mu_{ij}^2}{\varrho_{ij}}, & (\mathbf{r}_i - \mathbf{r}_j) \cdot (\mathbf{v}_i - \mathbf{v}_j) \leq 0 \\ 0, & \text{otherwise}, \end{cases} \qquad (21)$$

$$\mu_{ij} = \frac{h(\mathbf{v}_i - \mathbf{v}_j) \cdot (\mathbf{r}_i - \mathbf{r}_j)}{(\mathbf{r}_i - \mathbf{r}_j)^2 + \eta^2}, \qquad (22)$$

where $c_{ij}$ and $\varrho_{ij}$ are the arithmetic means of the sound velocity $c$ and the density $\varrho$, respectively. The parameter $\eta \approx 0.1h$ prevents the Monaghan–Gingold tensor $Q_{ij}$ from becoming infinite, if $|\mathbf{v}_i - \mathbf{v}_j| \neq 0$ and $r_{ij} = |\mathbf{r}_i - \mathbf{r}_j|$ tends to zero. $\alpha$ and $\beta$ are free parameters. Usually one sets $\alpha = 0.5$ and $\beta = 1$. For problems involving strong shocks the choice $\alpha = 1$ and $\beta = 2$ is more appropriate to avoid post–shock oscillations. The Monaghan–Gingold viscosity produces much more satisfactory results for shock waves and even strongly interacting shock waves can be simulated (Steinmetz & Müller 1993). Its disadvantage, however, is that it may be finite, although $\nabla \mathbf{v} = 0$. This typically happens in flows dominated by shear, i.e., where $\nabla \cdot \mathbf{v} = 0$, but $\nabla \times \mathbf{v} \neq 0$. Since neighboured particles have a different velocity, $\mu_{ij}$ and, therefore, $Q_{ij}$ is non zero, though the weighted average $\nabla v$ vanishes. Consequently, unwanted shear viscosity arises. For simulations of the formation of disk galaxies this spurious shear viscosity can be large enough to yield a substantial mass redistribution in the disk over a Hubble time, especially for small particle numbers ($<1\,000$). As a result, mass is steadily streaming towards the center of the galaxy.

These problems can almost completely be prevented by means of a modified artificial viscosity tensor $\widetilde{Q}_{ij}$ similar to that proposed by Balsara (1995):

$$\widetilde{Q}_{ij} = Q_{ij} \frac{f_i + f_j}{2}$$

$$f_i = \frac{|\langle \nabla \cdot \mathbf{v} \rangle_i|}{|\langle \nabla \cdot \mathbf{v} \rangle_i| + |\langle \nabla \times \mathbf{v} \rangle_i| + 0.0001 c_i/h_i}, \qquad (23)$$

where the term $0.0001 c_i/h_i$ is introduced to prevent divergences. In case of a shear-free, compressive flow, i.e., $\nabla \cdot \mathbf{v} \neq 0$, $\nabla \times \mathbf{v} = 0$ and, therefore, $f = 1$. The viscosity is identical to that proposed by Monaghan & Gingold (1983). However, in the presence of shear flows, the viscosity is diminished. The suppression factor $f$ is an order of magnitude estimate of the irrotational fraction of the flow. In case of a pure shear flow, i.e., $\nabla \cdot \mathbf{v} = 0$, $\nabla \times \mathbf{v} \neq 0$, $f$ vanishes and the viscosity is completely suppressed. Using (23), unphysical angular momentum transport can be reduced significantly. In case of the original Monaghan & Gingold viscosity the half angular momentum radius of a galactic size gaseous disk, which consists of 280 particles, grew by a factor of 2 within 3 Gyr. After the modification, the half angular momentum radius varied by less than 10% within a Hubble time (Steinmetz & Navarro in preparation).

The intrinsic assumption of smooth fields and the use of an artifical viscosity makes SPH not especially well suited to handle problems with strong shock waves. Numerical test, however, show that SPH gives surprisingly good results in predicting the correct states to the left and to the right of the discontinuity, even for small particle numbers (Steinmetz & Müller 1993). The accuracy with which the post shock state is calculated can compete with that obtained in most finite difference methods, though SPH is clearly inferior to shock capturing schemes using Riemann solvers like the *piecewise* *parabolic method* (PPM, Colella & Woodward 1984). The discontinuity itself is smoothed over a few smoothing length $h$. Since, however, in typical galaxy formation simulations the smoothing length $h$ is by a factor of several smaller than the grid spacing in simulations using finite difference methods, the resolution of shock fronts may still be superior. A probably more serious problem is that due to viscosity gas in front of the shock is preheated (Shapiro *et al* 1995). This effect may be unproblematic for pre shock gas. However, since in the absence of any shock wave the thermal energy of the gas can be much smaller than its kinetic energy, the numerical noise may be sufficient to give rise to a negative $\nabla \mathbf{v}$, i.e., the gas is artificially preheated to temperatures of several $10^4$ K and subsequently begins to cool (see below). Further work is required in order to clarify whether this effect flaws current galaxy formation simulations.

*3.4. Variable smoothing length*

In the derivations above, I assumed that all particles have the same smoothing length $h$ and that $h$ is time independent. However, both assumptions can be given up. In fact, SPH became most popular after people decided to incorporate an individual smoothing length $h_i$ for every particle $i$. Adapting the smoothing length to the local particle density, allows one to achieve a much higher spatial resolution in regions of high particle density. There are two different philosophies to interpret the smoothing procedure as described above. In the gather formulation, $A(\mathbf{r}_i)$ is determined by averaging over all particles within a sphere of radius $h_i$, i.e.,

$$A_i = \sum \frac{m_j \, A_j}{\varrho_j} W(r_{ij}, h_i) \,. \tag{24}$$

In the scatter formulation, $A(\mathbf{r}_i)$ results from an average over all particles $j$ which overlap particle $i$ with their sphere of radius $h_j$, i.e.,

$$A_i = \sum \frac{m_j \, A_j}{\varrho_j} W(r_{ij}, h_j) \tag{25}$$

(for a discussion, see Hernquist & Katz 1989). If one requires, that the interparticle force is antisymmetric, one has to use a combination of both. As in the case of the equation of motion there does not exist an unique way to antisymmetrize the force. Using the derivation starting from the least action principle a formulation arises in which always the sum $\frac{1}{2}(W(r_{ij}, h_i) + W(r_{ij}, h_j))$ appears. Another widely used formulation symmetrizes the smoothing length $h_{ij} = \frac{1}{2}(h_i + h_j)$ (Benz 1990).

Assuming a smoothing length which varies in space and time, one has to include terms $\propto \nabla h$ and $\propto \frac{dh}{dt}$ in the equation of motions. These terms are, however, difficult to estimate and so far most authors simply neglect them (see, however Nelson & Papaloizou 1993), although it may cause severe non–conservation of entropy/energy (Hernquist 1993). Further work is required in order to systematically analyze the effect of $\nabla h$ and $\frac{dh}{dt}$ terms.

Another, more technical problem arising from a variable softening length is the determination of $h_i$. The natural choice is to adapt it to the local density $h \propto (m_i/\varrho_i)^{1/3}$. The proportionality constant should be sufficiently large so that enough particles contribute to the sum in (12). On the other hand it should be as small as possible in order to save computing time. A widely used compromise is to choose $h$ so large that about 40 neighbours contribute to (12). A stability analysis by Morris

(1994), however, shows that this number may be too small to guarantee dynamical stability. A choice of about 60 neighbours seems to be more appropriate. In a first approach, Miyama, Hayashi & Narita (1984) adapted the softening at every time step using the density from the previous time step. This simple approach, however, often turns out to be dynamically unstable. Hernquist and Katz (1989) fixed the smoothing length by requiring every particle having a fixed number of neighbours (typically 40 neighbours which contribute to (12)). At every time step, the smoothing length $h_i$ is iteratively determined. This scheme, has the big advantage that the number of neighbours is fixed and the neighbour list can be kept in main memory. However, it can also have a pathological degeneracy, although it may be encountered only rarely. Consider a dense knot of gas consisting of $N_n - 1$ particles in an environment of diffuse gas, $N_n$ being the constant number of neighbours which contribute to (12). Suppose now than by chance, a particle from the environment approaches the dense knot of gas. Then, the smoothing lengths of the gas particles in the gas knot will change according to the motion of only one particle, and the density of the gas knot scales with $(\Delta r)^{-3}$, $\Delta r$ being the distance from the the environmental particle to the gas knot. Physically, this scaling seems to be not very plausible. In the approach of Benz (1990), Evrard (1988) or Steinmetz & Müller (1993), the smoothing is coupled to the gas density or a smoothed estimate of it, which almost excludes pathological situations as above. Due to the continuity equation a simple evolution equation for the smoothing length can be derived (Benz 1990): From $h \propto \varrho^{1/3}$ it follows

$$\frac{dh_i}{dt} = -\frac{1}{3}\frac{h_i}{\varrho_i}\frac{d\varrho_i}{dt} \tag{26}$$

$$= \frac{1}{3} h_i \nabla \cdot \mathbf{v}|_i \; . \tag{27}$$

The disadvantage of this approeach is that the number of neighbours can vary significantly, i.e., a minimum or maximum neighbour number cannot be guaranteed. This can be avoided by requiring that $h$ falls within a given interval of allowed numbers of neighbours (Steinmetz 1996). For some applications, these schemes may still be not stable enough: Small fluctuations in the density trigger fluctuations in $h$ which can amplify the density fluctuations and so on. We found that this behaviour can be avoided by relating $h$ not directly to $\varrho$ but to a smoothed estimated of $\varrho$. Such a smoothed estimated can easily be obtained by applying the SPH smoothing to $\varrho$ a second time, eventually even by using a different, somewhat larger smoothing length (Steinmetz & Müller 1993). Note, that for all hydrodynamic quantities the standard density is used, only in order to get a spatially smooth $h$ field, the doubly smoothed density is used.

A further generalization of SPH is to use anisotropic kernels, e.g., ellipsoids instead of spheres (Shapiro *et al* 1995, Owen *et al* 1995). These techniques were developed recently, but have not yet been applied to real simulations. The problem which arises with tensor smoothing is that the particle forces are no longer spherically symmetric, and, therefore, angular momentum conservation cannot be guaranteed. The advantage is that in sheetlike and filamentary structures, which are typical of galaxy formation simulations, a much higher spatial resolution can be achieved along the axis of maximum collapse.

*3.5. Time stepping*

Suppose a system at a time $t^n$. Position and velocities of a particle $i$ are denoted by $\mathbf{r}_i^n$ and $\mathbf{v}_i^n$, respectively. In the case of a gas particle, it also has an internal energy $\varepsilon_i^n$. Using the SPH formalism and Poisson's equation we can determine the acceleration $\frac{d}{dt}\mathbf{v}_i^n$ and the rate of energy change $\frac{d}{dt}\varepsilon_i^n$. In order to achieve a time integration which is at least second order accurate in time, we use the following predictor–corrector scheme:

(i) Predict $(\mathbf{r}, \mathbf{v}, \varepsilon)$ at time $t^{n+1} = t^n + \Delta t$ by means of a first order extrapolation: $\tilde{x}^{n+1} = x^n + \Delta t\, \dot{x}^n$, $x$ being an arbitrary element of the vector $(\mathbf{r}, \mathbf{v}, \varepsilon)$ which contains all positions, velocities and energies.

(ii) The density is calculated from equation (12). Together with the predicted energy $\tilde{\varepsilon}$, the equation of state is solved and the pressure $P$ is obtained.

(iii) Using the predicted positions and velocities as well as the updated from (ii), new accelerations and energy change rates are determined using equations (14) and (16).

(iv) Positions, energies and velocities are corrected by $x^{n+1} = x^n + \frac{1}{2}\left(\dot{x}^n + \dot{x}^{n+1}\right)$. The equation of state is solved again.

This scheme gives only a rough outline of a time integration scheme. Details can be found in Hernquist & Katz (1989) Monaghan (1992), Navarro & White (1993), and in Katz, Weinberg & Hernquist (1995).

People not familiar with numerical hydrodynamics may be surprised that only a low order time integrator is used. Usually, higher order methods (e.g., Runge–Kutta) provide the best performance for a given accuracy, because they allow the largest time step $\Delta t$. But for stability reasons the timestep in hydrodynamical methods has to satisfy the Courant–Friedrich–Levy time criterion.

$$\Delta t < \Delta t_{\text{CFL}} = \min_i \frac{h_i}{c_i + 0.6\,(\alpha c_i + \beta \max_j \mu_{ij})} \tag{28}$$

where $\alpha$ and $\beta$ are the parameters of the artificial viscosity (Monaghan 1992). The CFL condition often poses a more restrictive constraint than any accuracy criterium, i.e., the efficiency of higher order schemes cannot be exploited. Moreover, using approximative schemes like a treecode or P3M, the force fluctuates from time step to time step by a small amount due to a different tree construction or work distribution between PM and PP. Though these fluctuations does not critically influence the dynamical behaviour of the system, they also prevent high order methods achieving a large time step. Another time step constraint comes from the requirement that the force should not change too much from time step to time step:

$$\Delta t < \Delta t_{\text{f}} = \min_i \sqrt{h_i \left|\frac{d\mathbf{v}_i}{dt}\right|^{-1}} \tag{29}$$

Computer experiments show that a safe choice is $\Delta t = \min(0.3\Delta_{\text{CFL}}, 0.1\Delta_{\text{f}})$. Note that if cooling/heating processes are included the energy equation must be solved implicitly (Hernquist & Katz 1989).

Equation (29) also highlights an often neglected problem when comparing the CPU cost of gas dynamical simulations with that of $N$–body simulations. Suppose one wants to perform a simulation of a typical galaxy like our Milky Way, which (including the bulge) has a total mass of about $M = 10^{10}\,\mathrm{M}_\odot$ in the inner 1 kpc. If one wants to achieve a spatial resolution of $h = 1\,\mathrm{kpc}$, the corresponding time step is $\Delta t_\mathrm{f} \approx \sqrt{h^3/(GM)} \approx 4\,\mathrm{Myr}$. Combined with the factor 0.1 mentioned above, a simulation which extends over a Hubble time would require of about 25 000 time steps. Consequently, simply the number of time steps makes a hydrodynamical simulation at least 10 times more expensive than a comparable N–body simulation †. A similar number of time steps can also be derived from the CFL condition, if one takes into account that near the center of galaxies shocks are quite frequent and, therefore, the $\mu_{ij}$ term in the CFL condition dominates over the sound velocity $c_s$.

Studying highly collapsed systems like galaxies, it is advantageous to apply a multiple time step scheme (Hernquist & Katz 1989, Benz 1990). These schemes can easily and efficiently be combined with a treecode, while so far they were not applied to P3M codes. A very large gain in efficiency can be obtained with GRAPESPH (Steinmetz 1996), since a direct summation code has only a very small overhead per time step. In a multiple time step scheme, every particle's time step $\Delta t_i$ is calculated. A binary hierarchy of time bins is created $\Delta t^k = 2^k \Delta t^0, K = 0, 1, \ldots$. Every particle is propagated on the largest time bin $\Delta t^k$ which is still smaller than $\Delta t_i$. Positions and velocities of particles on larger time bins are interpolated. Therefore, in highly collapsed systems, only particles close to the center or those which are strongly shock heated must be propagated on the small time step, but for most particles the forces need not to be calculated so often. It turns out, that a multiple time step scheme often speeds up a simulation by a factor of ten, and sometimes even more.

### 4. Cooling and heating processes

The terms $\Lambda$ and $Q$ in the energy equation (8) have not been discussed so far. They describe "atomic" processes which influence the dynamics of the gas, namely radiative cooling, photoionization heating and feedback processes due to star formation. While the cooling processes are fairly well understood (at least as long as atomic cooling of gas of primordial composition is concerned), our understanding of star formation and related feedback processes is rather poor.

*4.1. Radiative cooling and photoionization*

The primary cooling processes for a gas with a density similar to the interstellar medium (ISM) are collisional. At low temperatures ($T \approx 10^4\,\mathrm{K}$), the collisional excitation cooling is dominant: a neutral H † collides with a free electron and the atom is excited. The excited hydrogen decays to the ground state emitting a photon, which escapes (the gas is assumed to be optically thin). Hence thermal energy of the gas is transferred to excitation energy of the atom. This energy is radiated away and, therefore, lost by the system. A similar, but less efficient process is collisional ionization followed by recombination. Both collisional processes have a dominant temperature dependence $\propto \exp(-\chi/(kT))$, $\chi$ being the excitation (or ionization)

---

† Of course, a N–body simulation with a resolution of 1 kpc would require the same number of time steps. However such a high resolution is seldom necessary, since dark haloes are much less concentrated than gaseous objects (see below). Therefore, N–body simulations typically have a factor of a few larger softening lengths $\epsilon$ than a similar size hydrodynamical simulation

† I use H as an example, but the process works analogously for any not completely ionized atom

energy of the atom, typically of the order of 10 eV. For metal lines it can be as high as 1 keV. Though this factor is highest at large temperatures, the cooling rate also depends on the fraction of neutral hydrogen (a completely ionized gas cannot be excited), which is highest at low temperatures. The combined effect turns out to be most efficient at energies of about $2\,10^4$ K for neutral hydrogen and $10^5$ K for single ionized helium, respectively (figure 2, upper left). In the case of a metal enriched gas, similar processes can efficiently act at temperatures between $10^5$ and $10^7$ K. At higher temperatures, where the gas is completely ionized, the dominant cooling process is bremsstrahlung. Since all these processes are collisional, i.e., they are caused by two body encounters, the rate of energy loss scales with the number density $n$ as $n^2$. Therefore, the combined cooling rate due to all these processes, $\Lambda(\varrho, T)$, is usually tabulated as $\Lambda/n_{\rm H}^2$, which is primarily a function of the temperature $T$ only.

High redshift observations ($z > 2$), e.g., the missing Gunn–Peterson effect (1965), infer that at these redshifts an UV background is present which is able to completely ionize hydrogen. The frequency dependence of the UV background flux is usually approximated by a single power law:

$$J(\nu) = J_{-21} \left(\frac{\nu}{\nu_{\rm H}}\right)^{-\alpha} 10^{-21}\,{\rm erg\,cm^{-2}\,Hz^{-1}\,sec^{-1}\,ster^{-1}}, \qquad (30)$$

$\nu_{\rm H}$ being the ionization frequency of hydrogen (13.6 eV). UV spectra as inferred from Quasars correspond to $\alpha = -1.5\ldots-1$, while the UV spectra of massive stars have a much softer frequency distribution ($\alpha = -5$). Note, however, that (30) is only a poor description of the actual frequency distribution of the UV background. However, the dynamics and cooling properties of the gas are mainly determined by the cross section averaged UV flux (Miralda-Escude *et al* 1995):

$$\bar{J} = \frac{\int {\rm d}\nu\,\nu^{-1} J(\nu) \sigma(\nu)}{\int {\rm d}\nu\,\nu^{-1} \sigma(\nu)} \qquad (31)$$

rather than by the detailed frequency distribution.

An UV background can strongly affect the cooling properties, especially if the abundance ratios are close to primordial: Due to the presence of the UV field, all gas at low densities ($n \lesssim 10^{-2}\,{\rm cm^{-3}}$, depending on $\bar{J}$) is completely ionized, even at temperatures of a few $10^4$ K. Therefore, the dominant cooling process, collisional excitation cooling, is completely suppressed (see figure 2, upper right), which diminishes the cooling capability of the gas at these temperatures by two orders of magnitude. Moreover, the residual energy of the photon ($\nu - \nu_{\rm H}$) preheats the gas to temperatures of the order of $3\,10^4$ K. Note, that compared to the cooling processes discussed in the last paragraph, photoionization involves one ion (and one photon) and, therefore, it is proportional to the gas density. Consequently, photoionization primarily affects the cooling properties of a low density gas, while the behaviour of a high density gas ($n \gtrsim 1\,{\rm cm^{-3}}$, depending on $\bar{J}$) is almost unchanged (figure 2, lower panels). The cooling function $\Lambda/n_{\rm H}^2$ is no longer density independent, because a mixture of $\propto \varrho$ and $\propto \varrho^2$ effects contribute to $\Lambda$. Due to the preheating and the strong suppression of gas cooling at low densities, Efstathiou (1992) argued that the formation of galaxies in low mass haloes with virial temperatures below several $10^4$ K can be suppressed. However, galaxies with such a small virial velocity preferentially form at high redshifts, where gas densities are higher and recombination ($\propto \varrho^2$) may dominate over photoionization ($\propto \varrho$). Whether an UV field is able to significantly affect masses of galaxies and on which mass scale it is important, is still a matter of debate (Cen & Ostriker 1993, Thoul & Weinberg 1995b, Steinmetz 1995, Quinn, Katz & Efstathiou 1996).

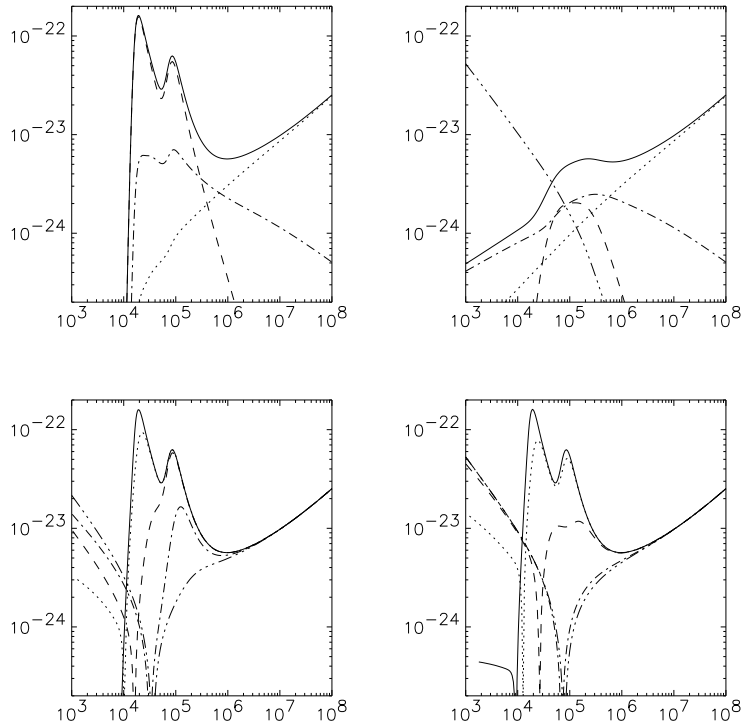

**Figure 2.** Upper left: cooling function $|\Lambda/n_H^2|$ for a gas of primordial composition without UV background (solid line) and physical processes which contribute: collisional excitation (dasehd), recombination (dashed–dotted), and bremsstrahlung (dotted). Upper right: processes contributing to the cooling function $|\Lambda/n_H^2|$ of a gas with $n_H = 10^{-3}\,\mathrm{cm}^{-3}$ in the presence of a photoionizing UV background field. , as well as photoionisation heating is shown by the dashed–triple–dotted line. Lower panels: Combined cooling function $\Lambda/n_H^2$ in the presence of an UV background field for different gas densities ($n_H = 10^{-6}\,\mathrm{cm}^{-3}$ (dashed-triple-dotted), $n_H = 10^{-4}\,\mathrm{cm}^{-3}$ (dashed-dotted) ,$n_H = 10^{-2}\,\mathrm{cm}^{-3}$ (dashed), $n_H = 1\,\mathrm{cm}^{-3}$ (dotted) and $n_H = 10^2\,\mathrm{cm}^{-3}$ (solid)). Left panel: Hard UV spectrum ($J_{-21} = 1, \alpha = 1$). Right panel: soft spectrum ($J_{-21} = 1, \alpha = 5$). The soft spectrum cannot completely suppress the He peak except for very low densities. Note, that at low temperatures, the gas is photo heated. The temperature where heating and cooling balances decreases with increasing density.

Another important cooling process is inverse Compton cooling with photons of the microwave background. If a low energy photon hits an electron, it is Thompson–scattered. On average the radiation field will gain energy since the radiation energy of the background is lower than the energy of the electrons. Consequently, the thermal energy of the gas is depleted. Since Compton cooling is proportional to the energy of the radiation field, it scales with $(T_{\mathrm{CMB}}(1+z))^4$. Due to the strong $z$ dependence, it is most important at high redshifts ($z > 5$), where the cooling time is smaller than the Hubble time. Note, that the Compton scattering process involves one electron and one photon, i.e., the cooling rate is $\propto \varrho$.

*4.2. Star formation and feedback processes*

Galaxies consist of stars and if (massive) stars are formed, a very strong energy source is provided: A high mass star ($> 8\,\mathrm{M}_\odot$) explodes as a Type II supernova and releases $10^{51}$ erg. However this energy is partially emitted as cosmic rays (see e.g., Dorfi 1990) and, therefore, is not immediately available to heat up the ISM. A similar amount of energy is released by stellar winds. In the following "supernova feedback" is used as a synonym for all feedback processes which are exerted by stars. As far as the global structure of a galaxy is concerned, a very simplified picture is justified, namely that over its lifetime every high mass star injects $10^{51}$ erg into the ISM with an efficiency of several percent. Due to the short lifetime of high mass stars ($10^7$ yr) compared to the dynamical time of a Galaxy (several $10^8$ yr), this energy is deposited into the ISM almost instantaneously. For a standard initial mass function (Miller & Scalo 1979), about 0.5% of all stars are high mass stars. For a galaxy like the Milky Way, where the total number of stars is about $10^{11}$, this implies an energy source of almost $10^{60}$ erg, which is higher than the total binding energy of the Galaxy, including a dark halo! Therefore, feedback processes due to star formation are likely to affect the dynamics of a galaxy strongly.

Unfortunately, the process of star formation itself and also how supernovae and stellar winds interact with the interstellar medium is only poorly understood. The best we can do so far (at least on the macroscopic scale of a galaxy) is to empirically parameterize the problem. Most models so far follow the implementation as outlined by Katz (1992): It is assumed that the star formation rate $\frac{d\varrho_*}{dt}$ is proportional to the local gas density divided by the local dynamical time scale

$$\frac{d\varrho_*}{dt} = c_* \frac{\varrho_{\mathrm{gas}}}{\tau} \qquad (32)$$

$c_*$ being the star formation efficiency, typically a factor of about a few per cent. Since the local dynamical time scale $\tau \propto 1/\sqrt{G\varrho}$, the star formation rate grows with the gas density like $\varrho^{1.5}$. Over a typical star formation time step $\Delta t$, a collisionless star particle (which represents of the order of a few million stars) of mass

$$m_* = m_{\mathrm{gas}} \left(1 - \exp\left(-\frac{c_* \Delta t}{\tau}\right)\right) \qquad (33)$$

is created and the mass of the gas particle is correspondingly reduced. Over the lifetime of a high mass star the supernovae energy is released and the corresponding mass and energy injected to the gas. According to the SPH formalism, energy and mass are distributed over a sphere with a radius corresponding to the smoothing length of a particle. In Steinmetz & Müller (1994, 1995) supernovae also metal enrich the ISM. So far simulations exhibit that the energy injected by supernovae affect the dynamics of a galaxy only little. Most of the energy is radiated away. Indeed, in Steinmetz & Müller (1995) it is shown, that the net energy loss, i.e., the difference between the energy injected by supernovae and the energy radiated away due to cooling is almost identical in simulations with and without supernova feedback. Navarro & White (1993) pointed out that the small influence of feedback is a result of the way it is implemented, namely that energy is added to the thermal energy of the gas. If energy is added to the kinetic energy, the effect of feedback is quite drastic. Even in the case of a low feedback efficiency of about 1%, the energy released by the first supernovae is able to strongly suppress further star formation in galaxies with virial velocities less than about 100 km/sec. In summary, the effect of star formation is still unclear since its influence strongly depends on the implementation and on the model parameters. Furthermore, none of the simulations so far was able to create a multi-phase medium like the ISM of our Milky Way (see, however, Klypin, this volume). Besides a probably too simplistic star formation model, the simulations also lack sufficient numerical resolution to give a realistic representation of the ISM. It is questionable whether the increased power of the next generations of computers will allow this resolution problem to be overcome.

## 5. Numerical simulations of galaxy formation

### 5.1. An analytic galaxy formation model

Based on the cooling properties of the gas and some heuristic assumptions on star formation and feedback processes, one is able to derive an analytic galaxy formation model. Its primary concept is that during the collapse of a galaxy the gas is heated up to virial temperature due to shock heating. Whether or not such a halo of hot gas is able to form a galaxy, depends on whether or not this gas is able to cool on a timescale short compared to its dynamical time scale. Following White (1995) the limiting mass (in gas), below which gas can rapidly cool, can be written as

$$M_{\text{gas}} = f\, M_{\text{halo}} = 3\ 10^{13}\, f^2\ \text{M}_\odot, \tag{34}$$

where $f$ is the fraction of baryons within a dark halo. With $f \approx 0.05$, as inferred from primordial nucleosynthesis (Walker *et al* 1991), a limiting gas mass of $7.5\ 10^{10}\ \text{M}_\odot$ results, which is in remarkably good agreement with the total stellar content of the most massive observed spiral galaxies.

If the gas is able to cool, the temperature, and, therefore, also the pressure is decreasing and the gas collapses. The density increases and cooling becomes even more efficient. Because of its high cooling capabilities the gas is virtually isothermal and the collapse cannot be stopped by pressure support. This runaway of cooling gas is usually labeled as the *cooling catastrophe*. The collapse in the rotational plane is stopped due to the angular momentum barrier. The gas settles into a rotationally supported thin disk. Details of this approach can be found in Binney (1977), Rees & Ostriker (1977), Silk (1977) White and Rees (1978) and Fall & Efstathiou (1980).

Although the predictions of the analytical model are in surprisingly good agreement with the properties of observed spiral galaxies, one should be carefully interpret these models: According to the picture of hierarchical galaxy formation dark haloes do not undergo a spherical collapse but they are built up by a series of mergers of already collapsed objects of smaller mass. Applying the cooling arguments to the progenitors, which due to their earlier formation epoch possesses a higher density, on average all gas within such a progenitor halo has already been cooled off before the haloes merge. Therefore, there is no diffuse gas left in the halo which could be shock heated to virial temperature. All gas within a halo should be cool gas and also the argument with the maximum mass (equation (34)) seems to be questionable. At least in galaxy clusters, however, we observe that most of the baryons are in form of hot and diffuse gas. Within the last few years the analytical approach has become more attractive (White & Frenk 1991, Kauffmann, Guiderdoni & White (1993), Lacey

& Cole 1993, Cole *et al* 1994) by combining it with the Press–Schechter algorithm (Press & Schechter 1974) and its extended version (Bower 1991, Bond *et al* 1991): For a given cosmogony ($\Omega_0$, $\Lambda$, power spectrum) and a given redshift, the Press–Schechter algorithm gives the number distribution of haloes as a function of their mass. Furthermore one can obtain information about the formation history of a halo, i.e., it is taken into account that major galaxies are build up by a serious of mergers. As argued above, these models confirm the cooling catastrophe. As an immediate consequence, it is predicted that every dark halo possesses a baryonic core and this core contains all the gas which originally belonged to the halo. Far too many low mass galaxies are formed (Kauffmann, Guiderdoni & White 1993). Therefore, feedback due to supernovae must be an important ingredient to every galaxy formation model. It regulates which haloes form stars and which fraction of the gas is converted into stars. In the phenomenological models described in this paragraph, star formation is implemented in a heuristic way involving two free parameters: the star formation parameter $\alpha$, which relates the star formation rate to the gas density and the feedback efficiency $\epsilon$, a parameter which determines how much of the energy injected by supernovae actually heats up the gas and how much is radiated away. Furthermore it is assumed that gas cooling and supernovae heating is in equilibrium. The free parameters $\alpha$ and $\epsilon$ are calibrated such that the luminosity and the gas content of a galaxy similar to our Milky Way is reproduced. This approach turned out to be a very powerful tool in order to study e.g., the global evolution of galaxy populations and how it depends on the background cosmogony. Also the influence of the environment (e.g., formation of field galaxies versus cluster galaxies) can be investigated.

Another powerful alternative to hydrodynamical simulations is to combine the analytic galaxy formation models as outlined in the last paragraph with the dynamics of $N$–body simulations (Kauffmann, Nusser & Steinmetz 1995). The advantages compared to hydrodynamical simulations are that (i) $N$–body simulations need fewer time steps and the computing time per time step is only a small fraction of that for hydrodynamical simulations. (ii) Since the detailed structure of the halo is less interesting (it is provided by the analytic models), even objects owning only a few particles can be analyzed, while objects whose dynamics is not disturbed by discreteness effects need to have quite a large particle number of at least 500. Therefore, the combined $N$–body plus analytic galaxy formation models allow one to statistically investigate synthetic galaxy catalogs which are much denser than those derived from comparable gas-dynamical simulations.

**Figure 3.** The distribution of gas projected in the X-Y and Y-Z plane shown for 6 different redshifts $z =$. The gas infall is mainly lumpy. Diffusely infalling gas settles down to form a rotationally supported disk.

*5.2. Results of numerical simulations*

Numerical simulations of forming galaxies is a challenging problem: The large collapse factors of the gas imply that galaxies are quite concentrated compared to their host dark haloes. The luminous matter of a typical Milky Way sized galaxy can be found within a disk with diameter of about 40 kpc. In order to numerically resolve the galaxy, the resolution must not be worse than a few kpc. In cosmological scenarios like the well known *Cold Dark Matter* scenario, such a galaxy is located in a halo of about 300 kpc radius. Even more, the halo acquires its mass from a region with a diameter of up to several Mpc. In order to describe the collapse history and the tidal field sufficiently accurate, large scale modes with wavelengths up to 40 Mpc or more should be included. In summary simulations are required whose dynamical range exceeds a factor of 10 000.

In the following, results of an ongoing project with Julio Navarro are presented (Steinmetz & Navarro, in preparation). In order to perform numerical simulations which exhibit a dynamic range of 30 000, we applied a multi-mass technique as described in Porter (1985), (see also Katz & White 1993; Navarro & White 1994; Bartelmann, Steinmetz & Weiss 1995): Based on a coarsely grained, large scale N-body simulation, single haloes are picked out. Every halo is then re-calculated in a high-resolution run including gas dynamics. The high resolution run uses the same large-scale density modes, but adds smaller-scale modes up to the Nyquist frequency of the refined grid. The mass distribution outside the refined region is approximated by a few thousand macro-particles, whose mass is increasing with distance.

As described above, the analytic models predict, that gas which cools within a dark halo settles to form a thin rotationally supported disk. The rotational properties of a halo are characterized by the spin parameter $\lambda = J\sqrt{E}/(GM^{2.5})$, $J$, $E$, and $M$ being the angular momentum, energy, and mass of the system, $G$ is the gravitational constant. It is commonly believed, that the angular momentum of galaxies stems from tidal torques due to surrounding matter (Hoyle 1949, Peebles 1969, White 1984). $N$-body simulations as well as an analysis by means of linear perturbation theory predict a typical spin parameter of about $\lambda = 0.05$ (figure 5, see also Barnes & Efstathiou 1987, Steinmetz & Bartelmann 1995), which implies that dark haloes are slow rotators. According to Fall & Efstathiou (1980), such an amount of initial angular momentum would be sufficient to explain the scale lengths of disk galaxies, if the specific angular momentum is conserved, i.e., the gas collapses axisymmetrically. However, as discussed above, galaxies are supposed to built up by merging of smaller structures. The gas is not accreted axisymmetrically but falls in in lumps (see figure 3); as a result angular momentum can be transported between gas and dark halo. Indeed, numerical simulations of Navarro & Benz (1991) show, that the angular momentum transport from the gas to the halo is much larger than predicted by dissipationless simulations. As a consequence, most of the gas ends up in a dense gas knot close to the center of the dark matter halo. In a recent, more systematic study, Navarro, Frenk & White (1995a) found, that the specific angular momentum of the gas in the central gas knot is only 20% of that of the dark halo, which implies that these disks should possess a much smaller scale length than observed. However, the numerical resolution of these simulations was not sufficient to investigate the inner structure of the central gas knots. The larger numerical resolution of the simulations of the current work ($m_{\rm gas} = 4.5\ 10^6\ M_\odot$, softening 1 kpc) shows that gas which falls in diffusely (and also less bound gas lumps which are tidally disrupted early on) settles down to form a disk. However, most of the gas lumps are sufficiently tightly bound to resist the tidal field of the halo much longer. Due to dynamical friction they spiral down to the center and deliver most of their angular momentum to the dark halo. Note that this is an immediate consequence of the cooling catastrophe and the neglect of feedback processes. Indeed, in the upper right frame of figure 5 one can see that almost all gas which belongs to the halo is cooled and collapsed to the central region.

Figure 4 shows the final distribution of gas and dark matter for a typical halo formed in one of the simulations. At first glance, it seems to be quite similar to observed galaxies (and also similar to the predictions of the analytical model): the model galaxy consists of an extended spheroidal halo with a radius of about 300 kpc. The gas is much more centrally condensed and has settled in a thin disk with a diameter of about 50 kpc. A more careful analysis, however, shows that more than 70% of the gas can be found in the central gas clump within a radius of 2 kpc. This is also shown in the lower left frame of figure 5: The half mass radius of the gas is by a factor of 5 smaller than predicted by the collapse model of Fall & Efstathiou (1980). Compared with the specific angular momentum ($J/M$) of the halo, that of the gas is smaller by a similar factor, i.e., the gas has lost almost all its angular momentum. In fact, the gas in the inner 1 kpc is almost non–rotating. Subtracting this mass from the disk, the specific angular momentum of the residual disk is comparable to that of the dark halo. Consequently, the specific angular momentum of all the gas, disk *and* dense gas knot, is a factor of about four smaller than that of the halo, in agreement with the simulations of Navarro, Frenk & White (1995a). Compared to observations, the numerical models predict the formation of disks, which have the right size, but far too much mass is acquired by the central "bulge".

**Figure 4.** Three dimensional view of the distribution of gas (top) and dark matter (bottom) in a Milky Way sized halo ($v_c = 220\,\mathrm{km/sec}$) for three different projections. The box has a side length of 50 kpc.

Up to now, there is no self–consistent high–resolution simulation which avoids the angular momentum problem. In order to solve it, one has to take care, that more gas is diffusely accreted. A variety of different physical processes which might be able to solve this problem are currently under discussion:

(i) One possibility is to change the merging history of the forming galaxy, e.g., by using a fluctuation spectrum with less power on small scales as predicted, for example, by a cosmogony with hot and cold dark matter. Changing the cosmological parameters $\Omega_0$, $\Lambda_0$ has probably rather little influence: Although the merging rate in the near past can be changed a lot, the merging history, expressed in terms of number of mergers and mass distribution of progenitors, is quite similar (Lacey & Cole, 1993).

(ii) A UV background as discussed in section 4.1 may suppress the formation of small structures ($v_c \lesssim 50\,\mathrm{km/sec}$, Efstathiou 1992) and the gas might fall in more diffusely. Therefore, the angular momentum transport to the halo might be reduced. However, the simulation presented here exhibits only little influence of the UV background: Most of the central gas clump has formed at sufficiently high redshifts (i.e., at sufficiently high densities) that recombination ($\propto \varrho^2$) dominates over photoionization ($\propto \varrho$) (Steinmetz 1995). However, the background field is able to prevent cooling of the late and diffusely infalling gas, which possesses a low density. Indeed, on average the mass of the cooled object in a simulation with an UV background is 30% smaller than in a corresponding run without an UV field. However, the mass of the gas knot is almost identical while the disk is almost completely suppressed.

(iii) As discussed in section 4.2, a realistic galaxy formation model has to include the effects of star formation, and the solution of the angular momentum problem is most likely related to supernovae feedback. In a first approach we performed a simulation including star formation as described in 4.2. In this model supernovae increase the thermal energy of the surrounding gas. As discussed above, most of the energy is radiated away due to the high cooling capability of the gas. As a result, the formation of small lumps of gas is not prevented. The main influence of star formation is to transform a dense knot of gas into a slightly more diffuse lump of stars, but the extensive transport of angular momentum to the dark halo is not overcome. It is conceivable that momentum input due to supernovae might have a much stronger effect (Navarro & White 1993).

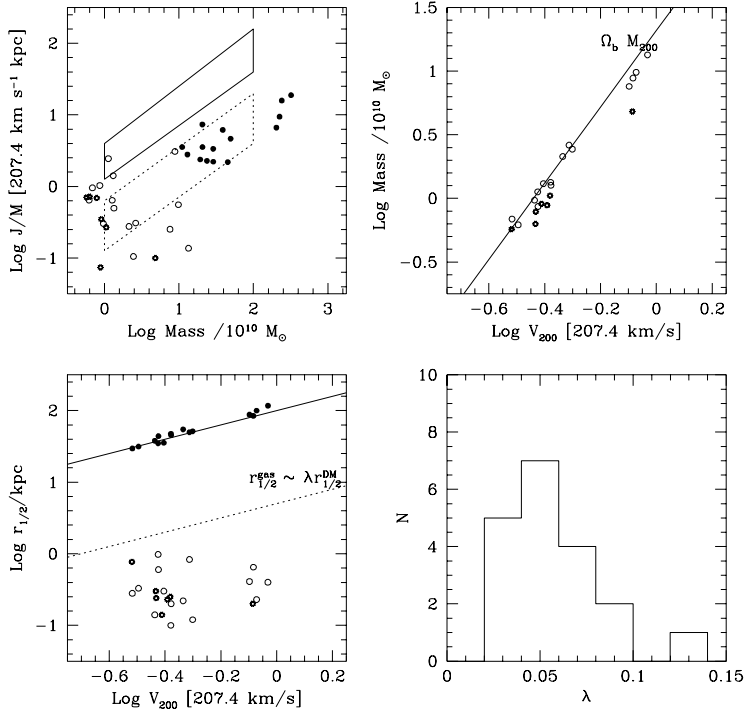

**Figure 5.** Upper left: specific angular momentum J/M versus mass for dark matter (filled circles), gas without UV background (open circles) and gas in a simulation with an UV background (starred symbols). The solid and dotted frame correspond to the observed J/M versus M distribution of spiral and elliptical galaxies, respectively. In case of an axisymmetric collapse, the points for the gas should have the same J/M, but a factor 20 less mass ($\Omega_b = 0.05$), i.e., the open and starred symbols should be just to the left of the filled circles. Since, however, the open circles and starred symples are to the lower left of the filled circles, this plot demonstrates that the gas has substantially lost angular momentum. Upper right: gas mass in the central disk (20 kpc) versus circular velocity of the halo. The solid line indicates the total gas mass within the halo. Since all symbols lie close to the solid line, almost all gas of the halo can be found in the central disk. In case of an UV background, the gas mass in the central object is suppressed by about 30%. Lower left: half mass radius versus circular velocity of the halo. It visualizes that the gaseous objects are very concentrated. In fact they are a factor of 5 more compact than predicted by the axisymmetric collapse model of Fall & Efstathiou (1980). Photoionization does not seem to have a major impact. Lower right: histograms of the spin parameter distribution of the halos. The spin parameters of the selected haloes are quite typical and do not exhibit systematically too low an angular momentum.

I want to end this lecture with a warning, that insufficient numerical resolution may mimic a feedback process: Due to the finite resolution, the maximum possible density which the gas can reach, and consequently the cooling capabilities of the gas are limited. Therefore, gas which falls into a dark halo cannot cool in a low resolution simulation while it would cool in a simulation with higher resolution. Insufficient numerical resolution, therefore, works like feedback due to star formation: if the density is higher than a given threshold, stars are formed and the feedback due to supernovae prevents further cooling and further collapse. In both cases gas is prevented from collapsing to densities higher than a given threshold. Finding a convenient and reliable measure of whether or not the numerical resolution is sufficient, is difficult, problem dependent and often not possible. Numerical simulations must, therefore, be carefully analyzed in order to separate physical effects from numerical imperfections. In fact, analyzing a numerical simulation is often much more time consuming than performing it, even if the CPU time per simulation is measured in weeks or months. On the other hand, the virtue of performing numerical simulations is to analyze systems, which are too complex in order to be immediately understandable by simple analytical arguments. It is the possibility to understand complex systems which makes numerical simulations so fascinating, but it is also this complexity which leads to the danger of being misled by misinterpreting numerical artefacts.

## Acknowledgments

I would like to thank Joel Primack for inviting me to Varenna. My collaborator, Julio Navarro, is gratefully acknowledged for the permission to discuss results of of our joint work in this lecture.